# A Symmetric Unified Transport and Charge Model for Metal-Oxide-Semiconductor Field-Effect Transistor from Diffusive to Ballistic Regimes

Chien-Ting Tung[1]
[1)] Synopsys, Sunnyvale, CA 94085 USA
(*Electronic mail: cttung.research@gmail.com)

***Abstract*— This paper presents a symmetric unified transport (UT) compact model for metal-oxide-semiconductor field-effect transistors (MOSFETs) that bridges drift-diffusion (DD) and ballistic transport (BT) regimes. The proposed model self-consistently accounts for both current and charge across the DD-BT transition. Quantum capacitance and carrier transport are incorporated into the charge density formulation. Drain-side velocity saturation and the source-side thermal velocity limit are unified within a single framework using a physically motivated high-field scattering length, enabling accurate modeling from DD square-law behavior to the ballistic limit. In addition, a physical channel charge and capacitance model is developed to capture capacitance reduction in the quasi-ballistic regime, which is not considered in standard compact models. The model is verified using theoretical analysis and experimental data from MOSFETs with multiple channel lengths, achieving accurate fitting using only physically motivated model parameters. The formulation is continuous and symmetric, and it passes both DC and AC symmetry tests.**

## I. Introduction

MOSFETs are the fundamental building blocks for modern integrated circuits (ICs). Today's ICs may contain MOSFETs with channel lengths ranging from the micrometer scale for analog design to the nanometer scale for digital design. As the channel length decreases, carrier transport gradually transitions from drift–diffusion (DD) to ballistic transport (BT). Advanced MOSFET technologies, such as FinFET, FDSOI, and GAA, can achieve channel lengths in the sub-20-nm regime, reaching more than 80% of the nondegenerate thermal velocity with a ballistic ratio of 0.5 – 0.7 [1], [2], [3]. Therefore, it is important to develop a physical compact model that can smoothly describe transport mechanisms across a wide range of channel lengths.

Researchers have been studying the unified model including DD and BT. The virtual-source (VS) model is a widely recognized framework for describing nanotransistors based on a scattering theory [4], [5], [6], [7]. It provides an intuitive explanation of quasi-ballistic transport by introducing apparent mobility through Matthiessen's rule, which connects ballistic and diffusive mobility. However, the basic VS model cannot accurately model long-channel MOSFETs exhibiting square-law behavior. To address this limitation, a modified VS emission–diffusion (ED) model was proposed to extend the VS formulation to long-channel devices [8]. The VSED model employs a transport-dependent critical length to describe both long-channel and short-channel devices within the VS framework. Nevertheless, it is not compatible with the well-established DD models used in industry-standard compact models, making integration into existing frameworks difficult.

Khandelwal *et al.* proposed a unified compact model bridging DD and BT based on the Berkeley Short-Channel IGFET Model (BSIM) [9]. In their approach, Matthiessen's rule is used to combine the DD velocity with the thermal velocity, enabling BT effects to be incorporated into the standard DD model. However, concerns remain regarding the physical interpretation of this formulation. The DD model is reformulated such that the current is expressed as the product of the average charge and the DD velocity, which is then combined with the thermal velocity to determine the final carrier velocity. Since the thermal velocity fundamentally limits the velocity of source charge rather than the average charge, this formulation is less physically sound.

Another approach is the distributed channel model, which divides the channel into multiple ballistic segments [10]. As the number of segments increases, the MOSFET behavior transitions from BT to DD. To reduce simulation runtime, a series connection of DD and BT MOSFET models has been proposed to reduce the number of ballistic segments. However, this approach still increases simulation time due to the introduction of additional internal nodes.

Despite these efforts, charge modeling under quasi-ballistic transport has been seldom discussed. Wei *et al.* demonstrated, using a simplified potential profile, that the total channel charge under BT conditions is smaller than that under DD conditions [11]. A TCAD study also showed this result [2]. However, a symmetric and unified charge model spanning DD to BT is still lacking.

In this paper, a unified transport (UT) model is proposed. Both current–voltage (IV) and charge–voltage (QV) models are developed to smoothly transition from DD to BT. The proposed model is physically sound and avoids the use of channel segmentation. It is continuous and symmetric, overcoming the limitations of conventional source-referenced quasi-ballistic models. The model is verified using MOSFET data with various channel lengths.

## II. Unified Transport Model

The proposed MOSFET model consists of three components: A) the charge density model, B) the current model, and C) the total charge model.

### A. Charge Density Model

A simple charge density model is chosen based on 1D Poisson and charge sheet approximation [12], [13]. The framework is generic, independent of the choice of model, and can be applied to more sophisticated charge calculations [14].

Two types of charge density models are considered in this work: the charge at the DD limit and the top-of-the-barrier charge model. Under DD conditions, the channel is near equilibrium, allowing carriers to relax to a local quasi-equilibrium state. Using the gradual channel approximation (GCA), the charge density is expressed as (1), where $V_{gi}$ is the intrinsic gate voltage, $V_c$ is the channel quasi-Fermi potential, and $q_i$ is the mobile charge. The threshold voltage $V_T$ is given by (1b), accounting for drain-induced barrier lowering (DIBL), where $V_{dsx}$ is the effective drain–source voltage and $\eta$ is the DIBL factor. $\eta$ is treated as a constant here for simplicity but it should be length dependent due to short channel effects (SCE). $V_{dsx}$ is a smoothed absolute $V_{dsi}$ as (1c) to ensure symmetry where $V_{dsi}$ is the intrinsic drain-source voltage and $\delta$ is the smoothing factor. $C_{ins}$ denotes the gate insulator capacitance, and $C_{eff}$ is the effective gate capacitance including quantum correction, which is assumed to follow (1d) with $C_Q$ as the quantum capacitance. The thermal voltage is denoted by $\phi_t$, and the subthreshold slope (SS) factor $n$ is computed by (1e) to account for short-channel effects, where $n_0$ is the nominal SS factor and $n_d$ is the voltage-dependent coefficient. $\delta$ is set to $\phi_t$ in this paper.

$$V_{gi} - V_c - V_T = \frac{q_i}{C_{eff}} + n\phi_t ln\left(\frac{q_i}{nC_{ins}\phi_t}\right) \tag{1a}$$

$$V_T = V_{T0} - \eta V_{dsx} \tag{1b}$$

$$V_{dsx} = \sqrt{{V_{dsi}}^2 + \delta^2} - \delta \tag{1c}$$

$$C_{eff} = ({C_{ins}}^{-1} + {C_Q}^{-1})^{-1} \tag{1d}$$

$$n = n_0 + n_d V_{dsx} \tag{1e}$$

The above equations assume GCA and are not applicable under BT [11]. In the ballistic regime, the channel is highly nonequilibrium, and local quasi-equilibrium cannot be assumed. Instead of using (1) to compute the mobile charge, the VS or top-of-the-barrier charge ($q_{\text{top}}$) can be calculated using the Landauer approach [6], [7]. By combining the Landauer approach and Poisson's equation, $q_{top}$ can be solved using (2). $V_{top}$ is the channel Fermi potential at the virtual source (VS), computed from (2b), with $V_{di}$ and $V_{si}$ denoting the intrinsic drain and source voltages. This equation evaluates the minimum of $V_{di}$ and $V_{si}$. Eq. (2a) accounts for carrier injection from both the source and drain with a transmission rate given by $T = \lambda/(\lambda + L)$, where $\lambda$ is the scattering length and $L$ is the channel length. More rigorously, $L$ should be replaced by the voltage-dependent critical length $\ell$ [5], [6], [8]. For simplicity, $L$ is used in this work, and the results show that this assumption does not affect the outcome because $T$ is not directly used to define the injection velocity of this model as the other models do [6], [8].

$$V_{gi} - V_{top} - V_T = \frac{q_{top}}{C_{eff}} + n\phi_t ln\left(\frac{2q_{top}}{nC_{ins}\phi_t(2 - T + Texp(-V_{dsx}/\phi_t))}\right) \tag{2a}$$

$$V_{top} = 0.5(V_{di} + V_{si} - V_{dsx}) \tag{2b}$$

### *B. Current Model*

Similar to the core model, the current model is also divided into two parts: the DD velocity ($v_{dd}$) and the BT velocity ($v_{bt}$). Then, based on the Lundstrom Antoniadis model [5], under the form of transmission rate of $\lambda/(\lambda + \ell)$, the effective injection velocity ($v_{eff}$) can be written as an inverse combination of $v_{dd}$ and $v_{bt}$ of Matthiessen's rule.

To obtain $v_{dd}$, the DD current density ($i_{dd}$) is first calculated. DD-based compact models have been widely studied and share a common form, as expressed in (3) [9], [12], [15], [16]. Here, $q_s$ and $q_d$ are the mobile charge densities at the source and drain end. $\mu_0$ is the low-field mobility defined as $\frac{v_{th0}\lambda}{2\phi_t}$ with $v_{th0}$ being the nondegenerate thermal velocity, and $U_A$ and $E_U$ are the mobility-degradation parameters. $\beta_1$ is the fitting parameter for velocity saturation, and $v_{sat}$ is the saturation velocity. Unlike many models that treat $v_{sat}$ as a constant, $v_{sat}$ in this work exhibits a length dependence following (3d). From the Monte Carlo simulation [17], it is observed that the drain velocity increases over the long channel limit as the channel length decreases, which is known as velocity overshoot. To model this phenomenon, a semi-physical model is proposed in [2]. It introduces an effective scattering length for saturation velocity ($\lambda_{sat}$) to model the reflection rate of the high-field scattering ($r = \frac{L}{\lambda_{sat}+L}$). Following the physical insight that drain velocity is not bound when no scattering exists and saturates to $v_{sat0}$ at long channel, $v_{sat}$ is modeled by $v_{sat0}/r$ as (3d) which resembles the empirical fitting function shown in [17]. Although not explicitly stated, the $|q_s - q_d|$ in (3a) is smoothed to ensure $C^\infty$ continuity.

$$i_{dd} = \frac{\mu_{eff}}{L} \frac{(q_s^2 - q_d^2)/(2C_{eff}) + n\phi_t(q_s - q_d)}{\left(1 + \left(|q_s - q_d|/(C_{eff}V_{dsat1})\right)^{\beta_1}\right)^{1/\beta_1}} \tag{3a}$$

$$\mu_{eff} = \frac{\mu_0}{1 + UA\left(q_{top}/(C_{eff}\phi_t)\right)^{EU}} \tag{3b}$$

$$V_{dsat1} = v_{sat} L/\mu_{eff} \tag{3c}$$

$$v_{sat} = v_{sat0}(1 + \frac{\lambda_{sat}}{L}) \tag{3d}$$

$q_s$ and $q_d$ are obtained by solving (1), which requires knowledge of the $V_c$. This model adopts a widely accepted approach to determine $V_c$ [15]. First, $V_c$ at the electrical source is assumed to be unaffected by transport and is equal to $V_{si}$ or $V_{di}$ depending on the voltage polarity. At the electrical drain, the lower bound of $V_c$ is determined using the saturation velocity $v_{sat}$ and current continuity. After accounting for MOSFET symmetry, the channel quasi-Fermi potentials at the source and drain ends can be expressed as (4). $q_{\text{top}x}$ denotes $q_{\text{top}}$ at the DD limit. To avoid recomputing (1), it is approximated by dividing $q_{\text{top}}$ by the transmission factor. $V_{\text{dsat2}}$ represents the maximum voltage across the channel, and (4d) and (4e) are used to smoothly clamp $V_C^{d,s}$ to $V_{\text{dsat2}}$ while accounting for voltage polarity, where the superscript denotes the channel position and $\beta_2$ is a fitting parameter. $V_C^{d,s}$ is fed into (1) to calculate $q_s$ and $q_d$.

$$q_{topx} = 2q_{top}/\left(2 - T + Texp(-V_{dsx}/\phi_t)\right) \tag{4a}$$

$$q_{min} = q_{topx} - C_{eff}V_{dsat1}\left(\sqrt{\frac{2q_{topx}}{C_{eff}V_{dsat1}}} - 1\right) \tag{4b}$$

$$V_{dsat2} = \frac{q_{topx} - q_{min}}{C_{eff}} + n\phi_t \ln\left(\frac{q_{topx}}{q_{min}}\right) \tag{4c}$$

$$V_{dsat}^{d,s} = V_{dsat2} + e^{\mp V_{dsi}/\phi_t} \tag{4d}$$

$$V_c^{d,s} = V_{si,di} \pm \frac{V_{dsi}}{\left(1 + \left((V_{dsx} + \delta)/V_{dsat}^{d,s}\right)^{\beta_2}\right)^{1/\beta_2}} \tag{4e}$$

By dividing $i_{dd}$ by $q_{\text{top}}$, the DD velocity is obtained as $v_{dd} = i_{dd}/q_{\text{top}}$. It is worth noting that $v_{dd}$ represents the carrier velocity at the virtual source, rather than the velocity evaluated at the average channel charge $q_{\text{avg}} = (q_s + q_d)/2$ as assumed in [9], [18]. The injection velocity is defined at the top of the barrier. Therefore, the physically correct approach is to apply Matthiessen's rule to combine $v_{bt}$ and $v_{dd}$, both defined at the VS, to obtain $v_{\text{eff}}$. Another issue with using $q_{\text{avg}}$ in the final current expression is that $q_{\text{avg}}$ is not well defined under BT. Although $q_s$ can still be evaluated under ballistic conditions as $q_{\text{top}}$, $q_d$ can no longer be obtained from (1) due to the breakdown of the GCA [11].

The expression for $v_{bt}$ is derived next. Since quasi-ballistic effects are captured by Matthiessen's rule, only the fully ballistic case is considered. The ballistic current density ($i_{bt}$) is expressed in (5) [7], where $v_{th}$ denotes the thermal injection velocity accounting for carrier degeneracy as modeled in (5b), with $\alpha$ and $m$ as fitting parameters.

$$i_{bt} = q_{top} v_{th} tanh\left(\frac{V_{dsi}}{2\phi_t}\right) \quad (5a)$$

$$v_{th} = v_{th0}(1 + \alpha(\frac{q_{top}}{C_{eff}\phi_t})^m) \quad (5b)$$

Equation (5a) can be approximated and rewritten as (6a) using $\tanh(x) \approx x/(1 + x^{\beta_3})^{1/\beta_3}$, which closely resembles (3a). $\beta_3$ is a fitting parameter and is set to 4 in this paper, closely matching tanh. In this form, by defining $v_{bt}$ as (6b), (6a) behaves as a saturation function that clamps $v_{bt}$ at $v_{th}$.

$$i_{bt} = q_{top} \frac{\frac{v_{th} V_{dsi}}{2\phi_t}}{\left(1+(\frac{v_{th}(V_{dsx}+\delta)}{2\phi_t v_{th}})^{\beta_3}\right)^{1/\beta_3}} \quad (6a)$$

$$v_{bt} = \frac{v_{th} V_{dsi}}{2\phi_t} \quad (6b)$$

Finally, the drain-source current $I_{ds}$ is calculated using (7) by replacing $v_{bt}$ in (6) with $v_{\text{eff}}$, where $W$ is the channel width. $v_{\text{eff}}$ is obtained using Matthiessen's rule as in (7b), and the thermal injection limit is enforced through the smoothed clamping functions in (7a). The $|v_{eff}|$ in (7a) is also smoothed to ensure $C^{\infty}$ continuity. Using this formulation, the critical length $l$ need not be determined without a detailed band-structure description, as the carrier velocity smoothly transitions between the ballistic and DD limits via Matthiessen's rule. In addition, the model incorporates both drain-side velocity saturation and the source-side thermal injection limit. One may notice that (7b) encounters a 0-divided-by-0 issue at $V_{dsi} = 0$. From both physics and mathematics (L'Hôpital's Rule), we can know that $v_{eff} = 0$ when $V_{dsi} = 0$ and the limits exist from both $V_{dsi} = 0^+$ and $V_{dsi} = 0^-$. Therefore, we can simply set the $v_{eff} = 0$ at $V_{dsi} = 0$ without worrying about the continuity which is a common practice in compact modeling.

$$I_{ds} = W q_{top} \frac{v_{eff}}{\left(1+(\frac{|v_{eff}|}{v_{th}})^{\beta_3}\right)^{1/\beta_3}} \quad (7a)$$

$$v_{eff} = \frac{v_{bt} v_{dd}}{v_{bt}+v_{dd}} \quad (7b)$$

### C. Total Charge Model

A compact model must capture not only IV but also QV/CV characteristics. Intrinsic gate, drain, and source charges are obtained by integrating the channel charge. Under the GCA and DD conditions, the channel charge distribution is well defined and can be integrated to yield total charges, as shown in [19]. However, this approach is not applicable in the ballistic regime, where the GCA breaks down and local equilibrium cannot be assumed. In this case, the charge distribution can instead be determined by using current continuity and energy conservation while assuming a channel potential profile [11].

To connect the two limits, a physical charge model is developed as follows. $q_{\text{top}}$ can be expressed as (8a), where $q_0$ is the nominal charge and $\Delta\phi$ is the potential deviation from the nominal value. Eq. (8a) can be reformulated as (8b) and separated into two terms. The first term represents the charge under local equilibrium, corresponding to the DD condition ($q_{\text{dd}}$), while the second term represents the charge in the ballistic limit ($q_{\text{bt}}$). Here, the channel charge is expressed as a weighted sum of $q_{\text{dd}}$ and $q_{\text{bt}}$. This result is applied to other positions along the channel without further proof. It should be noted that this approach differs from that in [20], which employs separate carrier transmission channels for the DD and BT components. In the proposed model, the current is calculated as a lumped quantity without separating the charges. The use of (8c) does not identify which portion of the charge experiences scattering and which does not. Instead, (8c) should be interpreted as indicating that some the carriers can relax to a local equilibrium state due to scattering. This model uses the concept of effective transmission rate, comparable to series resistance as defined by scattering theory, rather than employing separate transport channels for DD and BT as parallel resistors.

$$q_{top} = q_0 e^{\frac{\Delta\phi}{\phi_t}} \frac{2-T+Te^{-V_{dsx}/\phi_t}}{2} \quad (8a)$$

$$q_0 e^{\frac{\Delta\phi}{\phi_t}} \frac{2-2T+T+Te^{-V_{dsx}/\phi_t}}{2} = q_0 e^{\frac{\Delta\phi}{\phi_t}} \left[(1-T) + \frac{T(1+e^{-V_{dsx}/\phi_t})}{2}\right] \quad (8b)$$

$$q_{top} = (1-T) q_{dd} + T q_{bt} \quad (8c)$$

Therefore, the source- and drain-side charges under UT, $q_{s,d}^{ut}$, are expressed in (9a), where $q_{s,d}$ denotes the local-equilibrium charge obtained from (1) and (4). The superscript indicates the unified transport (ut) or the ballistic transport (bt). The subscript means the position in the channel. The ballistic charge, $q_{s,d}^{bt}$, is estimated using the method in [11]. By energy conservation, we can obtain the velocity of carrier at each channel position as (9b) where $v_{x0}$ is the velocity at the virtual source, $V(x)$ is the potential drop in the channel, $q$ is the elementary charge, and $m^*$ is the effective mass. Using (9b), current continuity and assuming linear potential, the ballistic mobile charge in the channel $q_i^{bt}(x)$ can be expressed as (9c). The parameter $\gamma$ is related to $v_{x0}$ and $m^*$, and is treated here as a fitting parameter. Based on (9c), $q_{s,d}^{ut}$ can be modeled by (9d & e) where $V_{dsi}^{\mp}$ is the one-sided drain–source voltage that smoothly clamps $V_{dsi}$ to its positive or negative component by (9f). Using that, $q_s^{bt}$ ($q_d^{bt}$) becomes the $q_{top}/\sqrt{1 + \gamma V_{dsi}}$ when $V_{dsi}$<0 ($V_{dsi}$>0). Because carrier acceleration is not significant until $V_{dsat}$, $V_{dsi}^{\mp}$ is smoothed using $V_{dsat}$ to mimic this effect as (9g), where $V_{dsat}$ is defined as a weighted sum of the DD $V_{dsat2}$ and $2\phi_t$. $\sqrt{1 + 0.5\gamma V_{dsat}}$ is added to (9d & e) to make sure $q_{s,d}^{bt}$=$q_{top}$ at $V_{dsi}$=0.

$$q_{s,d}^{ut} = (1-T) q_{s,d} + T q_{s,d}^{bt} \quad (9a)$$

$$v_x = \sqrt{{v_{x0}}^2 + 2qV(x)/m^*} \tag{9b}$$

$$q_i^{bt}(x) = q_i^{bt}(0)/\sqrt{1 + \gamma V_{dsi} x/L} \tag{9c}$$

$$q_s^{bt} = \frac{q_{top}\sqrt{1+0.5\gamma V_{dsat}}}{\sqrt{1-\gamma V_{dsi}^{-}}} \tag{9d}$$

$$q_d^{bt} = \frac{q_{top}\sqrt{1+0.5\gamma V_{dsat}}}{\sqrt{1+\gamma V_{dsi}^{+}}} \tag{9e}$$

$$V_{dsi}^{\mp} = 0.5(V_{dsi} \mp \sqrt{V_{dsi}^2 + V_{dsat}^2}) \tag{9f}$$

$$V_{dsat} = (1 - T)V_{dsat2} + 2T\phi_t \tag{9g}$$

The total intrinsic gate and drain charges, ($Q_{Gi}$, $Q_{Di}$), are obtained by substituting $q_{s,d}^{ut}$ into the DD charge model in [19]. However, since that model is derived under GCA, corrections are required for ballistic and quasi-ballistic conditions. It is found that the charge distribution and total charge can be reduced by the factors given in (9). With this modification, the conventional DD charge model can be extended to a UT charge model that accounts for BT. The final expression is shown as (10) where $q_{ia} = 0.5(q_s^{ut} + q_d^{ut})$, $dq = q_s^{ut} - q_d^{ut}$, $K_q = 2(q_{ia} + C_{eff} n\phi_t)$, and $K_{bt} = \sqrt{1 + T^b a\gamma V_{dsx}}$.

$$Q_{Gi} = \frac{WL}{K_{bt}}(q_{ia} + \frac{dq^2}{6K_q}) \tag{10a}$$

$$Q_{Di} = -\frac{WL}{K_{bt}}(q_{ia} - \frac{dq}{6}(1 - \frac{dq}{K_q}(1 + \frac{dq}{5K_q}))) \tag{10b}$$

$$Q_{Si} = -Q_{Gi} - Q_{Di} \tag{10c}$$

$K_{bt}$ is a factor introduced to account for BT effects, where $V_{dsx}$ follows the same formulation as (1c), with $V_{dsat}$ used as the smoothing factor. $K_{bt}$ is based on $\sqrt{1 + \gamma V_{dsi}}$ from (9d & e), although the actual factor is position dependent as (9c). Thus, to integrate it with the DD model with minimal change, a fitting parameter $a$ is added to compensate for discrepancy by assuming that the charge distribution is uniformly scaled by $K_{bt}$. To consider the length dependency and ballistic ratio, $T^b$ is further added with $T$ as the transmission factor and $b$ is a fitting parameter. Both $a$ and $b$ are empirical parameters to incorporate BT effects in the DD charge model.

## III. VALIDATION

First, the model is validated against theoretical predictions by comparison with ideal DD and BT models. Fig. 1 to 3 shows the model simulations compared to the ideal drift-diffusion model and the ballistic model with no 2D electrostatics, SCE, including DIBL, $V_T$ roll-off, and SS degradation to isolate transport effects. For this study, $\lambda$=16 nm, $\lambda_{sat}$=10 nm, $\gamma$=30, $a$=0.28, and $b$=1 are assumed.

Fig. 1 verifies the model behavior in the long-channel limit by comparing the IdVd and $C_{ggi}$Vg (intrinsic gate capacitance) curves with those of the DD model. At L=2 μm, and the device operates in the fully diffusive regime. In this case, (7a) reduces to $v_{dd}$, and the channel charge fully relaxes to local equilibrium. Consequently, the UT model reproduces the classic square-law behavior, yielding the same IV and CV curves as the DD model.

When the channel length is reduced to 30 nm, $\lambda$ becomes comparable to $L$, and the MOSFET enters the quasi-ballistic regime, where carriers can travel beyond the long-channel saturation velocity. The simulated IV and CV characteristics from the DD, BT, and UT models are shown in Fig. 2. The DD model employs an apparent mobility but does not include the length-dependent $v_{sat}$, allowing the velocity overshoot effect captured by the UT model to be highlighted.

As shown in Fig. 2a, the current predicted by the DD model is limited by $v_{sat0}$, whereas the UT model exceeds this limit and approaches the ballistic regime due to the scattering-aware $v_{sat}$ formulation in (3d). Fig. 2b presents the corresponding CV results. In the DD model, the ratio between the high-$V_d$ and low-$V_d$ CV curves decrease with channel length as a result of velocity saturation. This behavior is no longer valid in the quasi-ballistic regime, where carriers can exceed the saturation velocity near the drain, leading to channel charge reduction [2]. With (9) and (10), the UT model produces CV characteristics that lie between the DD and BT limits.

At L=1 nm, the channel exhibits more than 95% ballisticity. The effective velocity $v_{eff}$ of the UT model is close to $v_{bt}$, indicating that the device operation is limited by source injection rather than drain-side velocity saturation. Fig. 3 shows the characteristic curves under this condition. Both the IV and CV characteristics predicted by the UT model closely approach the ballistic limit.

After the theoretical validation, the model is also verified with experimental data. Fig. 4 to 7 demonstrates the fitting results of ETSOI fabricated by the same IBM process technology from L=980 nm to 30 nm [8], [21]. For all IdVd curves, the $V_{gs}$ are biased from -0.2 to 0.5 V with 0.1 V-step. For all IdVg curves, $V_{ds}$ are biased at 0.01 V, 0.05 V and 1 V. A single set of universal transport parameters with constant $v_{th}$ is summarized in Table I, using a source/drain resistance of 125 Ω·μm. The injection velocity at 30 nm is calculated to be $8.6 \times 10^6$ cm/s with a ballistic ration of 0.75 that closely matches the values reported in [1], [3]. This shows that the proposed model can correctly predict the transport from diffusive to ballistic regime using a single set of physically meaningful transport parameters. Although this work focuses on length dependent transport physics and modeling of MOSFETs, electrostatic, DIBL and SCE parameters also have length dependency due to the 2D electrostatic effects. $V_{T0}$, $\eta$, $n_0$, and $n_d$ are subjected to the effect of $V_T$ roll-off, DIBL, SCE, and SS degradation. On the other hand, $\beta_2$ representing how the drain potential is saturated is also affected by the 2D electrostatics and is length dependent. The author follows the presentation in [8] to treat those electrostatic parameters as individual constants for different channel length, which are summarized in Table II. The length dependent models for those electrostatic parameters are widely studied by compact models like BSIM [22]. Those existing models can be applied to the parameters in this paper with minimal modification.

TABLE I

Universal fitting parameters for ETSOI

| $\lambda$ (nm) | $\lambda_{sat}$ (nm) | $v_{sat0}$ (cm/s) | $v_{th}$ (cm/s) | UA |
|---|---|---|---|---|
| 15.8 | 9 | 8.5E6 | 1.14E7 | 4.4E-5 |
| EU | $\beta_1$ | $\beta_3$ | $C_{ins}$ (F/m[2]) | $C_Q$ (F/m[2]) |
| 2.5 | 4 | 4 | 0.031 | 0.06 |

TABLE II
Individual fitting parameters for ETSOI

| | 30 nm | 40 nm | 50 nm | 980 nm |
|---|---|---|---|---|
| $V_{T0}$ (mV) | -150 | -150 | -143 | -120 |
| $\eta$ (mV/V) | 55 | 45 | 30 | 5 |
| $n_0$ | 1.5 | 1.35 | 1.3 | 1.12 |
| $n_d$ (1/V) | 0.25 | 0.2 | 0.1 | 0 |
| $\beta_2$ | 1 | 1.5 | 2 | 8 |

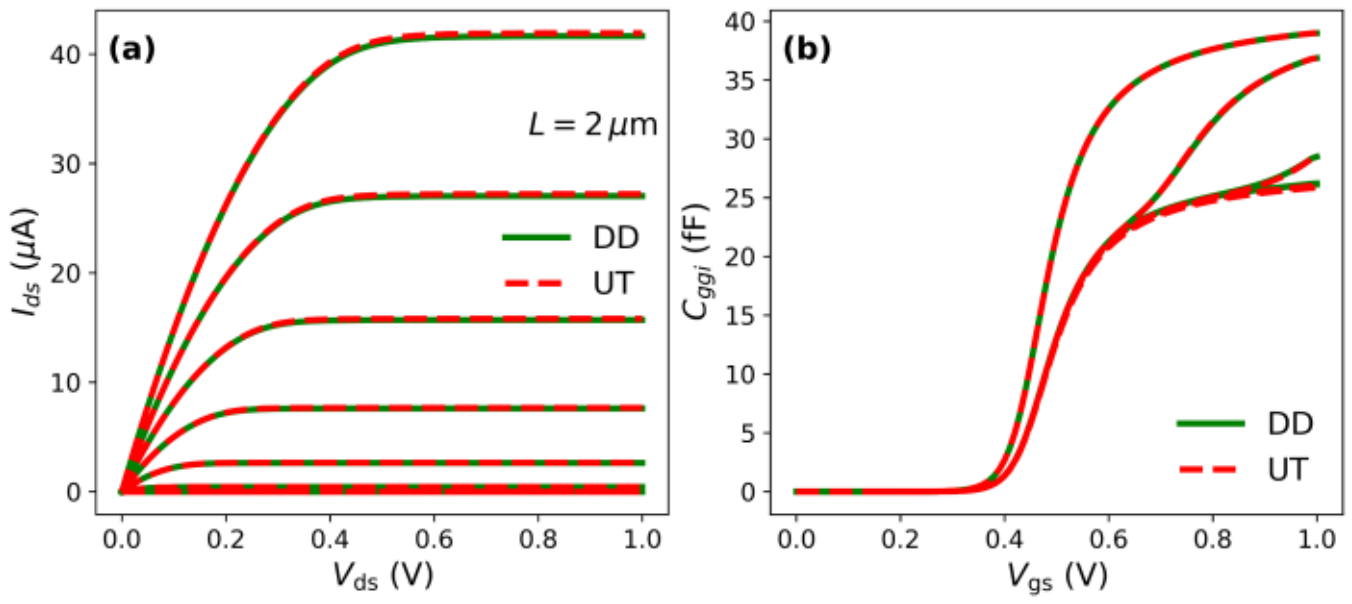

Fig. 1. (a) IdVd and (b) $C_{ggi}$Vg curves of the DD and UT model at L=2 μm.

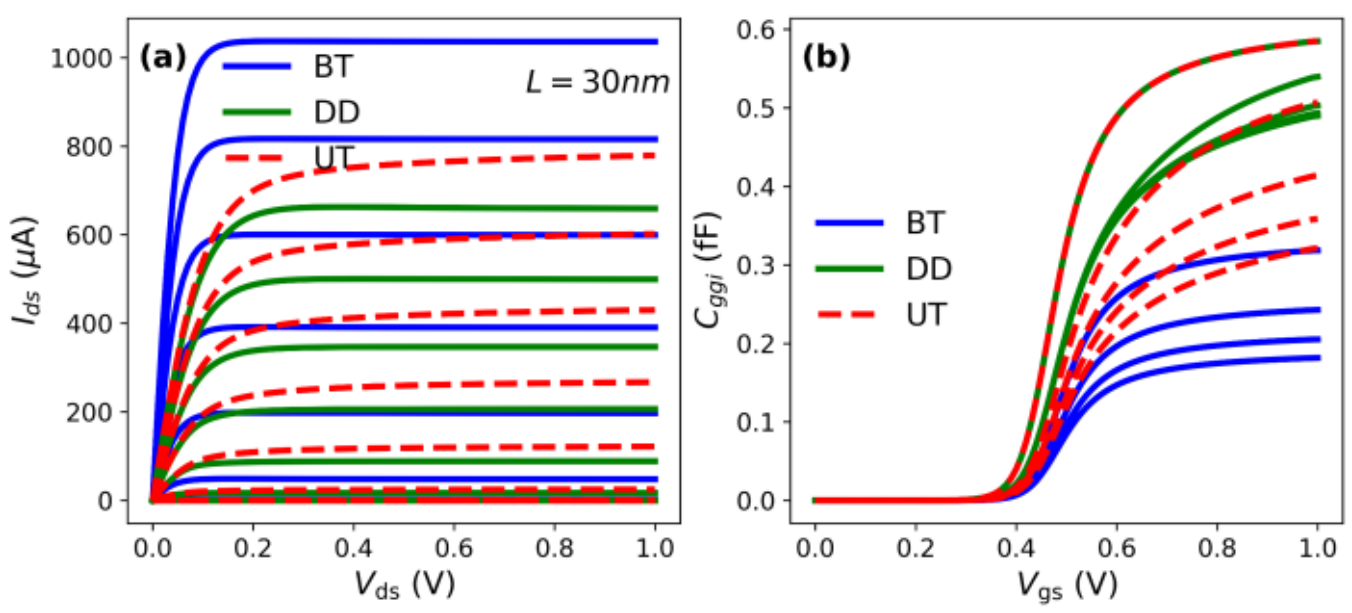

Fig. 2. (a) IdVd and (b) $C_{ggi}$Vg curves of the DD, BT and UT model at L=30 nm.

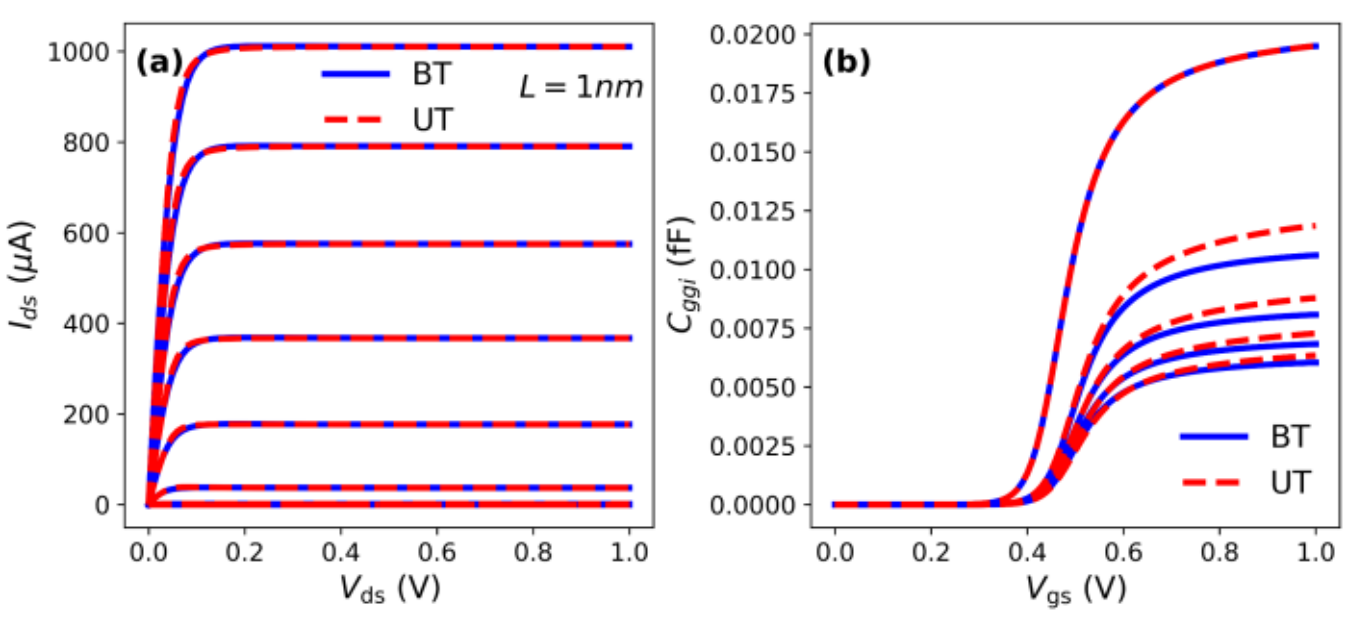

Fig. 3. (a) IdVd and (b) $C_{ggi}$Vg curves of the BT and UT model at L=1 nm.

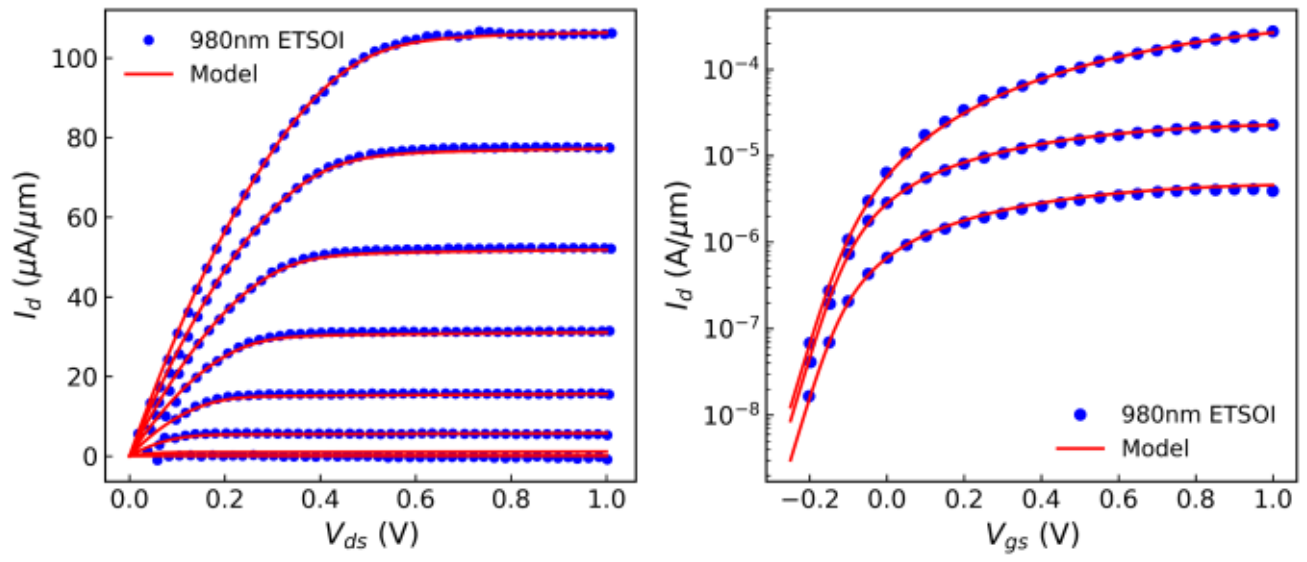

Fig. 4. (a) IdVd and (b) IdVg of a 980 nm ETSOI. Symbol: data [8]. Line: model.

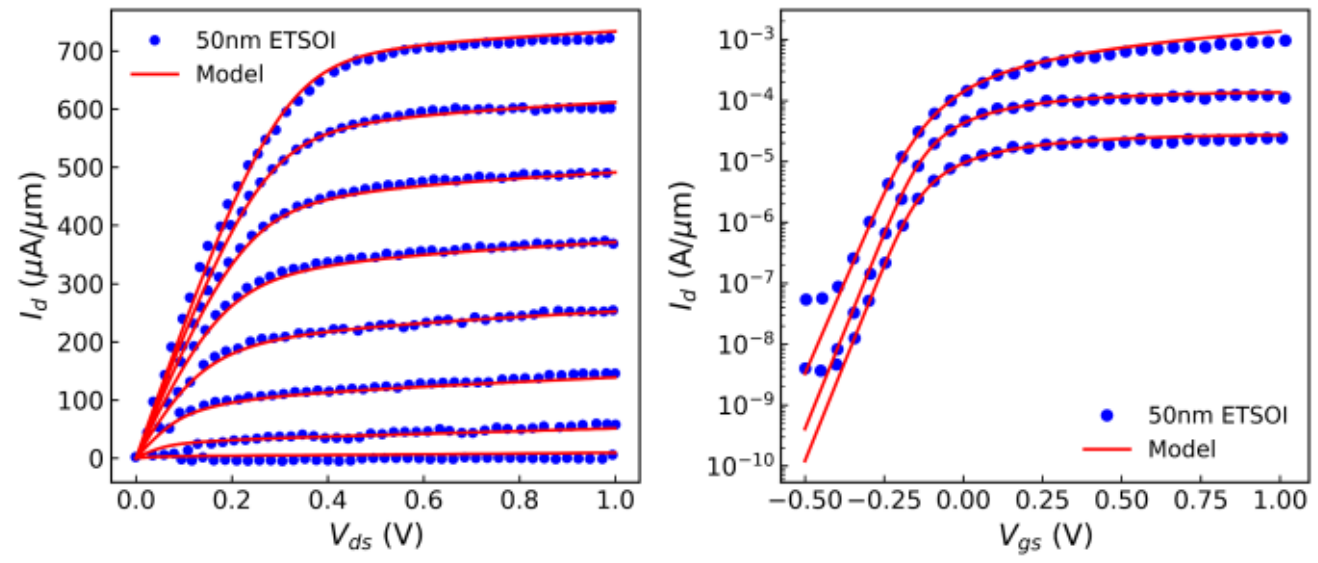

Fig. 5. (a) IdVd and (b) IdVg of a 50 nm ETSOI. Symbol: data [21]. Line: model.

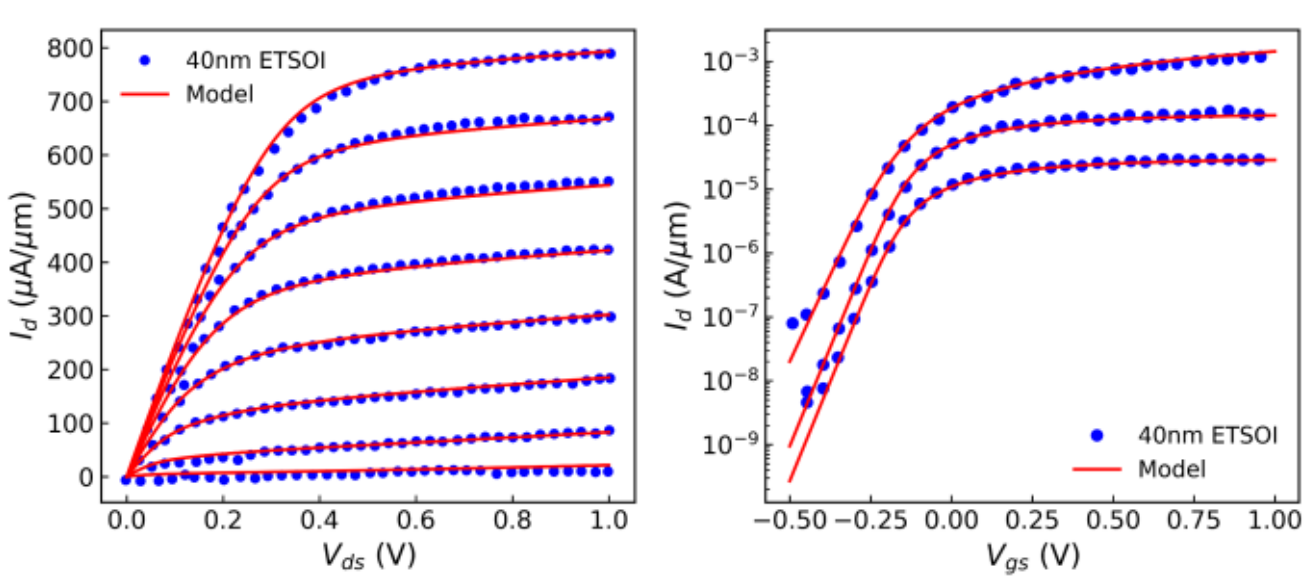

Fig. 6. (a) IdVd and (b) IdVg of a 40 nm ETSOI. Symbol: data [21]. Line: model.

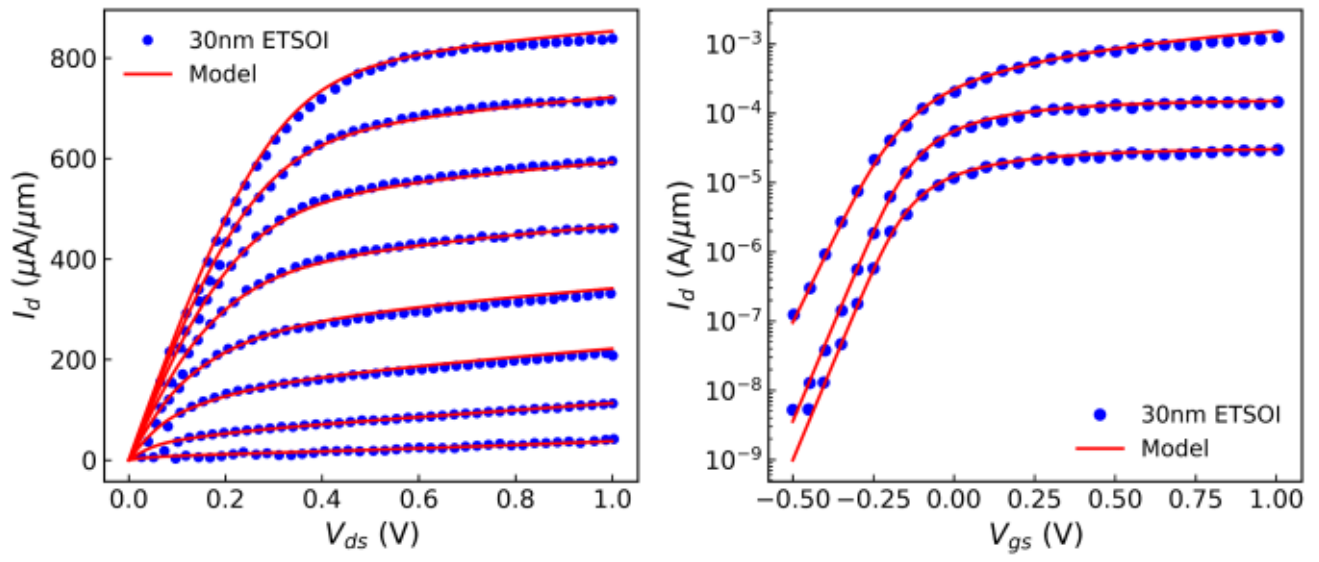

Fig. 7. (a) IdVd and (b) IdVg of a 30 nm ETSOI. Symbol: data [21]. Line: model.

Furthermore, the model is verified with an advanced FinFET [23], as shown in Fig. 8. Since the gate length and EOT are not reported in [23], a 16 nm channel length is chosen based on the technology node. By fitting the low $V_{DS}$ curve with $C_{eff}$=0.043 F/m$^2$, the inversion charge at $V_{GS}$=0.7 V is approximately 1.7 μC/cm². To achieve an on-current of 2300 μA/μm, an injection velocity of $1.35 \times 10^7$ cm/s is required, which exceeds the nondegenerate thermal velocity of silicon. Therefore, the degenerate $v_{\text{th}}$ model is used to capture this data, resulting in a maximum $v_{\text{th}}$ of $1.9 \times 10^7$ cm/s and a $v_{eff}$ of $1.44 \times 10^7$ cm/s from the model. The fitting parameters are summarized in Table III.

The model is developed with special attention to device symmetry. DC and AC symmetry tests are standard evaluations of MOSFET symmetry, smoothness, and continuity around $V_{ds}$=0 [24]. Most source-referenced models suffer from discontinuity at $V_{ds}$=0. Although the formulation of velocity saturation and ballistic transport inherently relies on source referencing due to the use of virtual-source charge for current and charge calculations, the proposed model avoids discontinuity and nonsmoothed behavior by introducing appropriate smoothing functions. As a result, the model successfully passes both DC and AC symmetry tests, as shown in Fig. 9, which is typically not satisfied by source-referenced quasi-ballistic transport models.

TABLE III
Fitting parameters for FinFET

| $\lambda$ (nm) | $\lambda_{sat}$ (nm) | $v_{sat0}$ (cm/s) | $v_{th0}$ (cm/s) |
|---|---|---|---|
| 10 | 10 | 1.8E7 | 1.14E7 |
| $\alpha$ | m | UA | EU |
| 0.3 | 0.3 | 7E-4 | 2.5 |
| $C_{eff}$ (F/m$^2$) | $V_{T0}$ (mV) | $\eta$ (mV/V) | $n_0$ |
| 0.043 | 270 | 22 | 1.12 |
| $n_d$ (1/V) | $\beta_1$ | $\beta_2$ | $\beta_3$ |
| 0.03 | 2.5 | 4 | 4 |

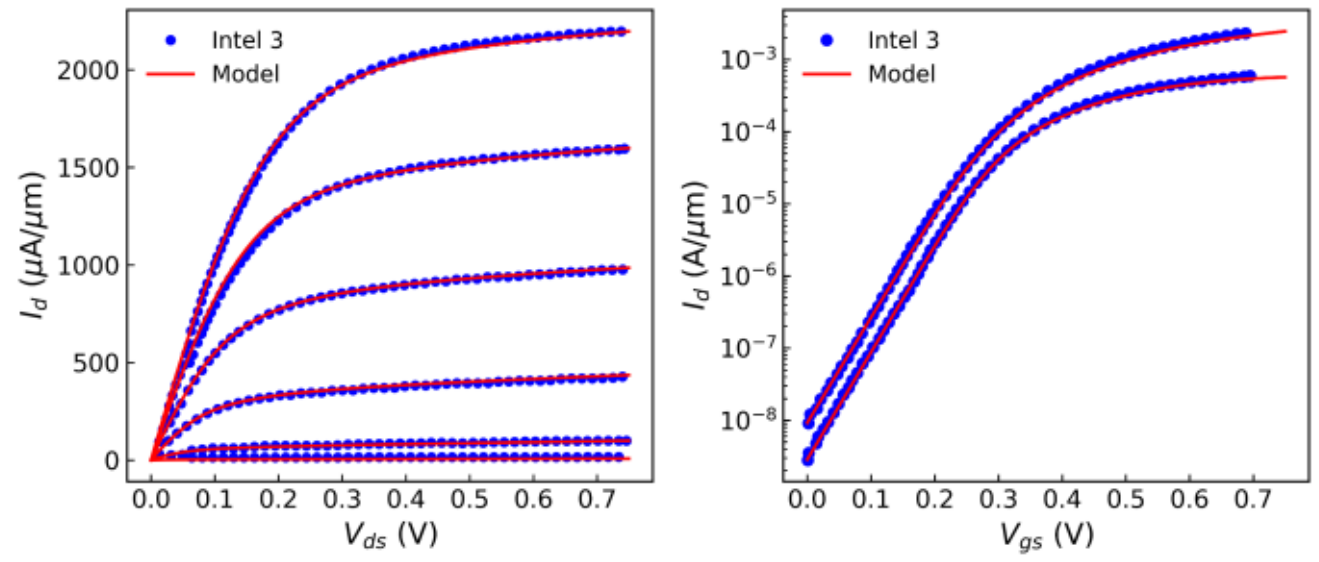

Fig. 8. (a) IdVd and (b) IdVg of an Intel 3 FinFET. Symbol: data [20]. Line: model. (a) $V_{gs}$ are biased from 0.2 to 0.7 V with 0.1 V-step. (b) $V_{ds}$ are biased at 0.05 V and 0.7 V.

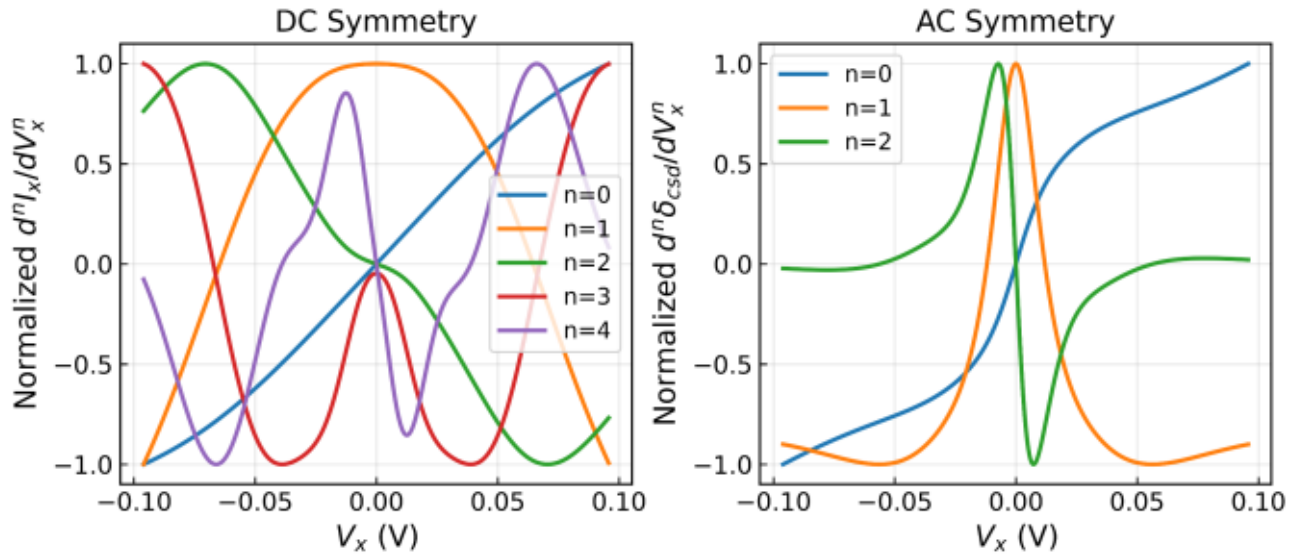

Fig. 9. The DC and AC symmetry test of this model.

## IV. Conclusion

A unified transport model for FETs is developed to describe carrier transport from the diffusive to the ballistic regime. The model improves conventional approaches for combining DD and BT velocities by introducing a physically consistent treatment of the thermal injection limit and velocity overshoot. A quasi-ballistic charge model is also derived to account for the transition between nonequilibrium and quasi-equilibrium charge distributions. The developed current and charge models are compatible with existing DD-based compact models with simple modifications. The model is verified using both theoretical and experimental data, demonstrating its predictive capability.

## Acknowledgement

The author would like to thank Prof. Chenming Hu for his mentorship, and Prof. Dimitri Antoniadis for his constructive and valuable feedback on this work.

This work was conducted independently and does not represent the views or intellectual property of Synopsys.